# Compensation of IQ-Imbalance and Phase Noise in MIMO-OFDM Systems

**Abstract**

The degrading effect of RF impairments on the performance of wireless communication systems is more pronounced in MIMO-OFDM transmission. Two of the most common impairments that significantly limit the performance of MIMO-OFDM transceivers are IQ-imbalance and phase noise. Low-complexity estimation and compensation techniques that can jointly remove the effect of these impairments are highly desirable. In this paper, we propose a simple joint estimation and compensation technique to estimate channel, phase noise and IQ-imbalance parameters in MIMO-OFDM systems under multipath slow fading channels. A subcarrier multiplexed preamble structure to estimate the channel and impairment parameters with minimum overhead is introduced and used in the estimation of IQ-imbalance parameters as well as the initial estimation of effective channel matrix including common phase error (CPE). We then use a novel tracking method based on the second order statistics of the inter-carrier interference (ICI) and noise to update the effective channel matrix throughout an OFDM frame. Simulation results for a variety of scenarios show that the proposed low-complexity estimation and compensation technique can efficiently improve the performance of MIMO-OFDM systems in terms of bit-error-rate (BER).

## I. Introduction

The reliability and throughput offered by orthogonal frequency division multiplexing (OFDM) and multi-input-multi-output (MIMO) techniques has made MIMO-OFDM a clear choice for future generation wireless communication systems [1]. However, MIMO-OFDM systems are highly sensitive to various radio frequency (RF) impairments due to the imperfections in analog front end components. Among various RF-impairments, IQ-imbalance and phase noise are considered more detrimental than others and can cause severe performance degradation in MIMO-OFDM systems.

Due to flexibility and low complexity, direct conversion and low intermediate frequency (IF) receivers are preferred to superheterodyne receivers. These receivers provide perfect image attenuation eliminating the need for image reject filters [2]. In practice, however, due to manufacturing imperfections, there is always phase and amplitude mismatch in the inphase (I) and quadrature (Q) branches of the IQ-receiver commonly known as IQ-imbalance [3]. The compensation of IQ-imbalance in MIMO-OFDM has gained a lot of attention recently [4], [5].

Phase noise is another major impairment caused by non-ideal oscillator which introduces random phase fluctuations at the oscillator output resulting in distortion of the signals [3]. The effect of phase noise is more severe in OFDM systems and is studied extensively in the literature [6], [7]. These works state that the effect of phase noise can be broken down into common phase error (CPE) and inter-carrier interference (ICI) with CPE being more prominent for systems with small phase noise. Performance analysis and compensation techniques are proposed for OFDM and MIMO-OFDM systems in the literature and can be found in [7]-[9].

Joint IQ-imbalance and phase noise compensation techniques for single-input-single-output (SISO) systems have been previously discussed in [10] and [11]. However, joint treatment of these impairments for MIMO-OFDM systems have not been studied extensively. To our knowledge, only recently in [12], a joint estimation and compensation scheme for MIMO-OFDM systems assuming quasi-static channel is proposed. The authors first develop a compensation scheme for SISO systems and generalize it to the MIMO case. The proposed method requires more training symbols with increasing number of transmit-receive antennas and the complexity of the scheme is high due to the required maximum likelihood (ML) estimation. In this paper, we first introduce a subcarrier multiplexed preamble structure to estimate the channel and impairment parameters with small overhead. The IQ-imbalance parameters as well as the initial estimation the effective channel matrix including common phase error (CPE) are then estimated on the preamble. Using scattered pilot symbols and second order statistics of ICI and noise, we then propose a novel tracking method based on the minimum mean square error (MMSE) to update the effective channel matrix in the rest of the OFDM frame. Simulation results for different phase noise and IQ-imbalance parameters show that the proposed low-complexity estimation and compensation technique can efficiently improve the performance of MIMO-OFDM systems in terms of bit-error-rate (BER).

Throughout the paper, upper-case bold letters are used for matrices, lower-case bold letters for vectors and scalars are represented by italic letters. Operators $\star$, $\odot$ and $\oslash$ indicate convolution, element-wise multiplication and element-wise division respectively. $\{.\}^*$, $\{.\}^T$, $\{.\}^H$, $\{.\}^\dagger$ and $\{.\}^\#$ represent conjugate, transpose, hermitian transpose, pseudo-inverse and conjugate mirror operation, respectively. $\mathbf{I}_N$, $\mathbf{0}_N$ represent $N \times N$ identity and zero matrix. $E\{.\}$, $\arg\{.\}$, $\text{abs}\{.\}$ represent expectation, argument and absolute value of the elements inside the parentheses, respectively.

## II. SYSTEM MODEL

### A. Signal Model

The block diagrams of typical MIMO-OFDM transmitter and receiver with $M_t$ and $M_r$ antennas are shown in Figs. 1 and 2, respectively. At transmitter, the data stream is first converted to parallel streams and passed through $N$-point IFFT block after adding pilot symbols. The resulting parallel streams are converted back to serial and a cyclic prefix (CP) of length $N_{cp}$ ($L \leq N_{cp}$; where $L$ is the number of resolvable channel paths) is then added to the signal stream. The resulting signal is up-converted with suitable modulation and transmitted. Let $\mathbf{s}_{p,m}$ be the $m^{th}$ $N \times 1$ frequency domain OFDM symbol to be transmitted from the $p^{th}$ transmit antenna whose time domain representation is given by $\tilde{\mathbf{s}}_{p,m} = \mathcal{F}^{-1}(\mathbf{s}_{p,m})$ where $\mathcal{F}^{-1}$ is the inverse Fourier transform operation. The vector that contains all the $N + N_{cp} + L - 1$ discrete time domain received signal samples at the $q^{th}$ receive antenna is then given by

$$\tilde{\mathbf{r}}_{q,m} = \sum_{p=1}^{M_t} \mathbf{h}_{q,p} \star \tilde{\mathbf{s}}_{p,m} + \mathbf{w}_q \qquad (1)$$

where $\mathbf{h}_{q,p} = (h_{q,p}(0), h_{q,p}(1), ..., h_{q,p}(L-1))^T$ is an $L \times 1$ vector representing the channel impulse response of the multipath channel from the $p^{th}$ transmit to the $q^{th}$ receive antenna for the $m^{th}$ OFDM symbol, $\tilde{\mathbf{s}}_{p,m}$ is the transmitted discrete time signal vector from the $p^{th}$ transmit antenna and $\mathbf{w}_q$ is an $(N + N_{cp} + L - 1) \times 1$ vector containing the samples of complex circularly symmetric additive white Gaussian noise (AWGN) at the $q^{th}$ receive antenna. At each receiver branch, the received signal is down-converted, sampled and have the cyclic prefix removed (see Fig. 2). The signal samples are then converted into parallel streams and fed into FFT block with size $N$. The equivalent frequency domain received signal model at the $k^{th}$ subcarrier ($k = 0, 1, .., N-1$) after FFT block can be written as

$$\mathbf{r}_m(k) = \mathbf{H}_m(k)\mathbf{s}_m(k) + \boldsymbol{\eta}_m(k) \qquad (2)$$

where $\mathbf{r}_m(k)$ is the $M_r \times 1$ received signal vector, $\mathbf{H}_m(k)$ is the $M_r \times M_t$ channel frequency response matrix, $\mathbf{s}_m(k)$ is the $M_t \times 1$ transmitted signal vector at the $k^{th}$ subcarrier of $m^{th}$ OFDM Symbol. $\boldsymbol{\eta}_m(k)$ is the $M_r \times 1$ frequency domain noise assumed to be complex zero mean AWGN with variance $\sigma_n^2$, i.e., $\boldsymbol{\eta}_m(k) \in \mathcal{CN}(0, \sigma_n^2)$. Each $(q,p)$ element of the channel frequency response matrix is given by $H_m^{q,p}(k) = \sum_{n=0}^{L-1} h_{q,p}(n) e^{-j\frac{2\pi k n}{N}}$.

### B. IQ-Imbalance Model

The time domain oscillator output with frequency independent IQ-imbalance (see Fig. 2) can be written as [3], [14]

$$\begin{aligned} c(t) &= \cos(\omega_c t) + j\varepsilon \sin(\omega_c t + \theta) \\ &= K_1 e^{j\omega_c t} + K_2 e^{-j\omega_c t} \end{aligned} \qquad (3)$$

where $\omega_c$, $\theta$ and $\varepsilon$ are the carrier frequency, phase mismatch and amplitude mismatch at the receiver, respectively, and $K_1 = (1 + \varepsilon e^{-j\theta})/2$, $K_2 = (1 - \varepsilon e^{j\theta})/2$. It is assumed that each receiver branch uses separate oscillators and IQ-imbalance parameters are different for each receiver branch. The received signal in frequency domain after FFT block can then be written as

$$\mathbf{x}_m(k) = \mathbf{K}_1 \mathbf{r}_m(k) + \mathbf{K}_2 \mathbf{r}_m^\#(k) + \boldsymbol{\eta}_m(k) \qquad (4)$$

where $\mathbf{x}_m(k)$ and $\boldsymbol{\eta}_m(k)$ are the $M_r \times 1$ received signal and noise vectors at $k^{th}$ subcarrier of the $m^{th}$ OFDM symbol, respectively, and $\#$ operation represents conjugate mirror operation i.e., $\mathbf{r}_m^\#(k) = \mathbf{r}_m^*(-k)$. The matrices $\mathbf{K}_1$ and $\mathbf{K}_2$ are given as $\mathbf{K}_1 = (\mathbf{I} + \boldsymbol{\varepsilon} e^{-j\boldsymbol{\theta}})/2$ and $\mathbf{K}_2 = (\mathbf{I} - \boldsymbol{\varepsilon} e^{+j\boldsymbol{\theta}})/2$ where $\boldsymbol{\varepsilon} = \{\varepsilon_i\}_{i=1}^{M_r}$ and $\boldsymbol{\theta} = \{\theta_i\}_{i=1}^{M_r}$ are the amplitude and phase mismatches associated with each of the receiver branches.

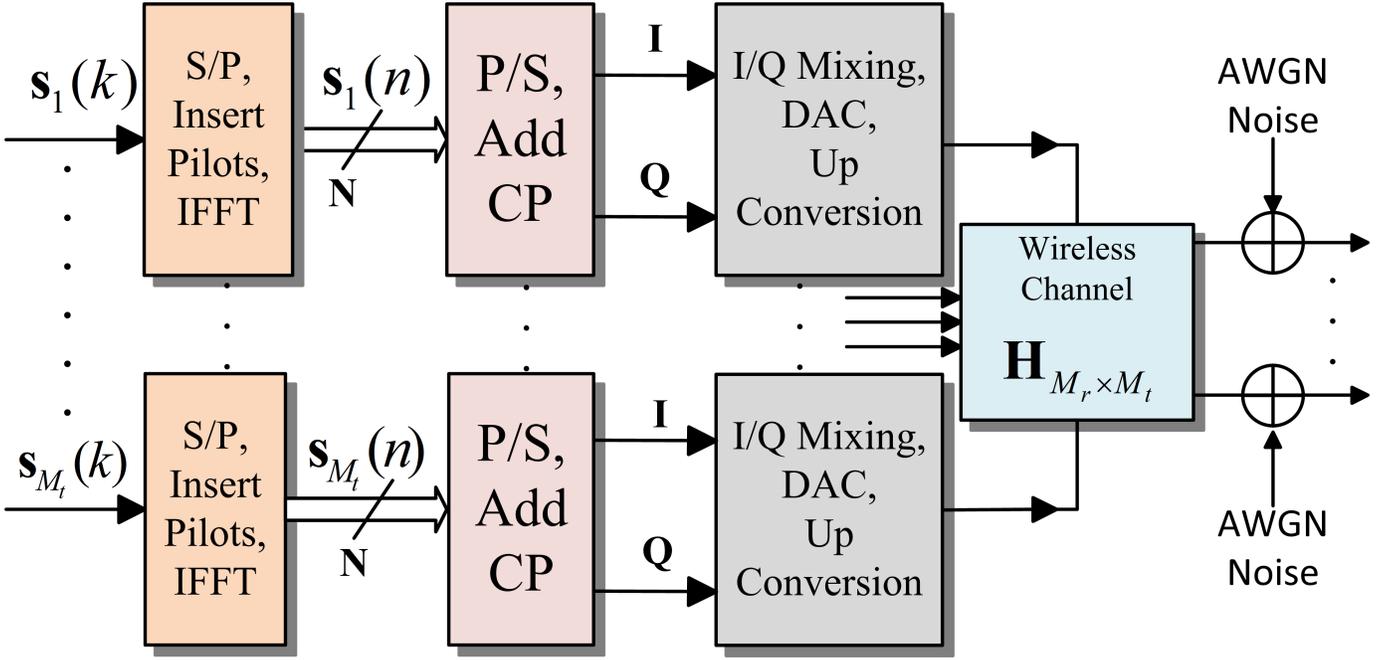

Fig. 1. MIMO-OFDM transmitter with $M_t$ transmit branches

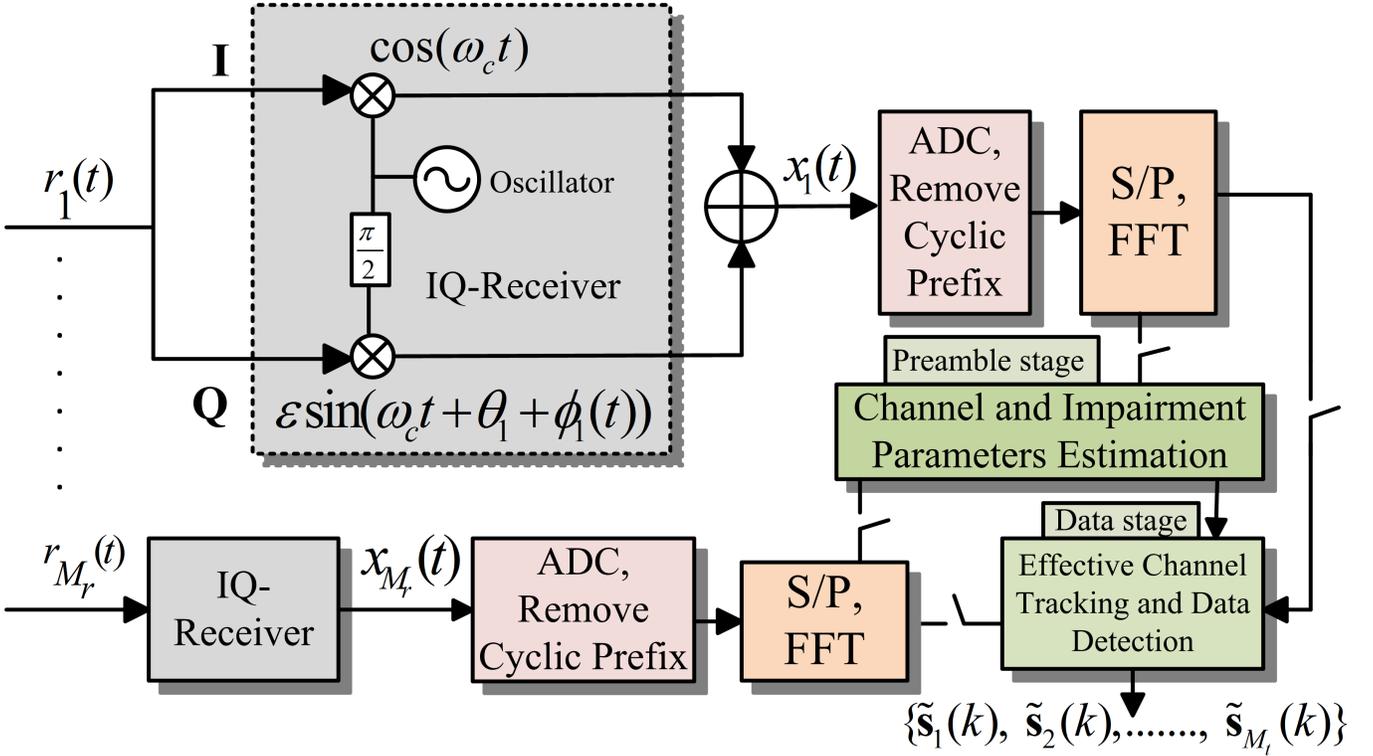

Fig. 2. MIMO-OFDM receiver with $M_r$ receive branches affected by IQ-imbalance and phase noise

### C. Phase Noise Model

In the presence of phase noise, the local oscillator output is given by $c(t) = e^{j(\omega_c t + \phi(t))}$ where $\phi(t)$ is the random phase fluctuations of the local oscillator. Phase noise process modeling has been extensively studied in the literature for both phase locked loop (PLL) based and free-running oscillators [7], [15]. Here, we consider a free-running oscillator model. The phase noise for this oscillator can be modeled as Wiener process with independent Gaussian increments such that $\phi(n+1) = \phi(n) + u(n)$, where $\phi(n)$ is the discrete Brownian phase noise process at $n^{th}$ time instant, $u(n) \in \mathcal{N}(0, 4\pi\beta T_s)$ with $\beta$ being the 3-dB linewidth of the phase noise spectrum and $T_s$ the sampling time period of an OFDM symbol. The time domain received signal corrupted by phase noise at the $q^{th}$ receiver can then be written as

$$\tilde{\mathbf{r}}_{q,m} = \left[e^{j\phi(0)}, e^{j\phi(1)}, ..., e^{j\phi(N+N_{cp}+L-1)}\right] \odot \sum_{p=1}^{M_t} \mathbf{h}_{q,p} \star \tilde{\mathbf{s}}_{p,m} + \mathbf{w}_q \quad (5)$$

where $e^{j\phi(n)}$ represents one realization of phase noise process for the $q^{th}$ receiver branch. Equivalently, the received signal in frequency domain can be written as

$$\mathbf{r}_m(k) = \sum_{i=0}^{N-1} \boldsymbol{\Theta}_m(i-k)\mathbf{H}_m(i)\mathbf{s}_m(i) + \boldsymbol{\eta}_m(k) \quad (6)$$

where $\boldsymbol{\Theta}_m(i)$ is the $M_r \times M_r$ diagonal matrix of phase noise frequency domain samples for the $i^{th}$ subcarrier ($i = 0, .., N-1$) of the $m^{th}$ OFDM symbol. The effect of phase noise for the received symbol can be separated into two terms, the CPE (for $i = k$) and ICI (for $i \neq k$) terms, as [7]

$$\mathbf{r}_m(k) = \underbrace{\boldsymbol{\Theta}_m(0)\mathbf{H}_m(k)\mathbf{s}_m(k)}_{CPE} + \underbrace{\boldsymbol{\zeta}(k)}_{ICI} + \boldsymbol{\eta}_m(k) \quad (7)$$

The first term in (7), $\boldsymbol{\Theta}_m(0)$, representing CPE is an $M_r \times M_r$ diagonal matrix showing the rotation of the received symbols by a common phase while the second term, $\boldsymbol{\zeta}(k)$, is the ICI term caused due to phase noise components from all other subcarriers at $k^{th}$ subcarrier given by $\boldsymbol{\zeta}(k) = \sum_{i=0, i \neq k}^{N-1} \boldsymbol{\Theta}_m(i-k)\mathbf{H}_m(i)\mathbf{s}_m(i)$. CPE is common for all subcarriers within one OFDM symbol but varies from one OFDM symbol to the other. Studies show that in practice CPE is more harmful to the system performance than ICI at realistic phase noise regimes where $\beta \ll 1/NT_s$ [7].

### D. Combined Phase Noise and IQ-imbalance Model

The combined effect of the phase noise and IQ-imbalance on the received symbol in frequency domain can be modeled by combining (4) and (7) as

$$\begin{aligned}\mathbf{x}_m(k) &= \mathbf{K}_1 \boldsymbol{\Theta}_m(0)\mathbf{H}_m(k)\mathbf{s}_m(k) \\ &+ \mathbf{K}_2 \boldsymbol{\Theta}_m^*(0)\mathbf{H}_m^{\#}(k)\mathbf{s}_m^{\#}(k) \\ &+ \boldsymbol{\zeta}(k) + \boldsymbol{\eta}_m(k)\end{aligned} \quad (8)$$

(8) reveals that the desired signal at particular subcarrier is affected by the joint effect of multiplicative and additive impairments due to IQ-imbalance and phase noise from all the subcarriers. We will show, in the results section, that a joint compensation of these impairments should be considered in order to achieve acceptable system performance.

## III. INITIAL CHANNEL AND IMPAIRMENT PARAMETERS ESTIMATION

### A. OFDM Frame Format

In typical MIMO-OFDM systems, such as those based on IEEE 802.11a/n standard, the transmitted OFDM symbols are divided into training and data symbols [16]. Training OFDM symbols are further classified into short training symbols and long training symbols also called as preamble. Furthermore, data type OFDM symbols are usually loaded with pilot symbols for fine tuning the estimation of channel and impairments. The null subcarriers in the short OFDM training symbol can also be used to estimate the noise plus ICI correlation matrix as given in [9]. For $L$ OFDM short symbols with $N_{null}$ null subcarriers, this correlation matrix $\boldsymbol{\Psi}$ can be estimated as [9]

$$\boldsymbol{\Psi} = \frac{1}{LN_{null}} \sum_{l=1}^{L} \sum_{k=1}^{N_{null}} \mathbf{x}_l(k)\mathbf{x}_l^H(k) \quad (9)$$

*B. Effective Channel and IQ-Imbalance Estimation*

The accuracy in estimating channel and impairment parameters at the preamble stage of OFDM transmission largely depends on the OFDM training (preamble) symbols design. Due to the presence of IQ-imbalance, interference from mirror subcarriers should also be considered in preamble design. On the other hand, utilizing more OFDM symbols for channel estimation at the preamble stage results in poor estimation performance due to different CPE at each training symbol [17]. In this paper, we use a modified subcarrier multiplexed preamble structure [18] with minimum overhead that is more robust against the varying nature of CPE. We define matrices $\mathbf{T}_1$ and $\mathbf{T}_2$ that contain $M_t$ training symbol vectors of size $N \times 1$ to be transmitted from $M_t$ transmit antennas as

$$\mathbf{T}_1 = \{\boldsymbol{\gamma} \circ \boldsymbol{\xi}_1, \boldsymbol{\gamma} \circ \boldsymbol{\xi}_2, ..., \boldsymbol{\gamma} \circ \boldsymbol{\xi}_{M_t}\} \tag{10}$$

$$\mathbf{T}_2 = \{\boldsymbol{\gamma} \circ \boldsymbol{\xi}'_1, \boldsymbol{\gamma} \circ \boldsymbol{\xi}'_2, ..., \boldsymbol{\gamma} \circ \boldsymbol{\xi}'_{M_t}\} \tag{11}$$

where $\boldsymbol{\gamma}$ is a $N \times 1$ vector of known training symbol that can be optimized according to the system requirements and the $N \times 1$ vectors $\boldsymbol{\xi}_p$ and $\boldsymbol{\xi}'_p$ corresponding to $p^{th}$ transmit antenna are given as

$$\boldsymbol{\xi}_p(k) = \Pi_{p-1}\{[1, \mathbf{0}_{M_t-1}, 1, \mathbf{0}_{M_t-1}, ..., 1, \mathbf{0}_{M_t-1}]\}^T \tag{12}$$

$$\boldsymbol{\xi}'_p(k) = \begin{cases} -\boldsymbol{\xi}_p(k) & \text{for } k \in \{1, 2, .., \frac{N}{2} - 1\} \\ \boldsymbol{\xi}_p(k) & \text{for } k \in \{-\frac{N}{2}, -\frac{N}{2} + 1, .., -1\} \end{cases} \tag{13}$$

In (12), $\Pi_{p-1}$ represents the cyclic shift operation over $p-1$ samples and $\mathbf{0}_{M_t-1}$ denotes an all-zero vector. Fig. 3 shows the preamble structure for the special case of 2 transmit antennas. With the above proposed preamble structure and without considering ICI and noise, the received signal vector for the two consecutive training symbols can be written as

$$\boldsymbol{\psi}_1(k) = \mathbf{K}_1 \hat{\mathbf{H}}_{pre}(k) \mathbf{t}_1(k) + \mathbf{K}_2 \hat{\mathbf{H}}^{\#}_{pre}(k) \mathbf{t}^{\#}_1(k) \tag{14}$$

$$\boldsymbol{\psi}_2(k) = \mathbf{K}_1 (\hat{\mathbf{H}}_{pre}(k) + \boldsymbol{\delta}) \mathbf{t}_2(k)$$
$$- \mathbf{K}_2 (\hat{\mathbf{H}}^{\#}_{pre}(k) + \boldsymbol{\delta}^*) \mathbf{t}^{\#}_2(k) \tag{15}$$

where $\mathbf{t}_1(k)$ and $\mathbf{t}_2(k)$ are the $k^{th}$ row of $\mathbf{T}_1$ and $\mathbf{T}_2$ with each of them containing only one nonzero element $\lambda_1(k)$ and $\lambda_2(k)$ respectively. $\hat{\mathbf{H}}_{pre}(k) = \boldsymbol{\Theta}_{pre}(0)\mathbf{H}(k)$ and $\hat{\mathbf{H}}^{\#}_{pre}(k) = \boldsymbol{\Theta}^*_{pre}(0)\mathbf{H}^{\#}(k)$ represent the effective channel matrix at the $k^{th}$ and $-k^{th}$ subcarriers with $\boldsymbol{\Theta}_{pre}(0)$ being the CPE associated with the training symbols. $\boldsymbol{\delta}$ is the $M_r \times 1$ vector representing error term caused due to different CPEs affecting two consecutive training symbols. This error term is the same for all the subcarriers. We define two vectors, $\boldsymbol{\chi}_a(k)$ and $\boldsymbol{\chi}_b(k)$ as

$$\boldsymbol{\chi}_a(k) = \{\boldsymbol{\psi}_1(k) + \boldsymbol{\psi}_2(k)\}/2\lambda_1(k) = \mathbf{K}_1 \hat{\mathbf{h}}_p(k) + \boldsymbol{\rho} \tag{16}$$

$$\boldsymbol{\chi}_b(k) = \{\boldsymbol{\psi}_1(k) - \boldsymbol{\psi}_2(k)\}/2\lambda^*_2(k) = \mathbf{K}_2 \hat{\mathbf{h}}^{\#}_p(k) - \boldsymbol{\rho} \tag{17}$$

where $\hat{\mathbf{h}}_p(k) = \boldsymbol{\Theta}_{pre}(0)\mathbf{h}_p(k)$ and $\hat{\mathbf{h}}^{\#}_p(k) = \boldsymbol{\Theta}^*_{pre}(0)\mathbf{h}^{\#}_p(k)$ represent the $p^{th}$ column vector of $\hat{\mathbf{H}}_{pre}(k)$ and $\hat{\mathbf{H}}^{\#}_{pre}(k)$, respectively and $\boldsymbol{\rho} = (\mathbf{K}_1\boldsymbol{\delta} - \mathbf{K}_2\boldsymbol{\delta}^*)/2$. Using (16) and (17) and the fact that $\mathbf{K}_2 = \mathbf{I}_{M_r} - \mathbf{K}^*_1$, we can approximate $\hat{\mathbf{h}}_p(k)$ as

$$\mathbf{e}(k) = \boldsymbol{\chi}_a(k) + \boldsymbol{\chi}^{\#}_b(k) = \hat{\mathbf{h}}_p(k) + \boldsymbol{\rho} - \boldsymbol{\rho}^* \tag{18}$$

From (18), the channel vector corresponding to the $k^{th}$ subcarrier ($k = kM_t + p, k = 0, 1, ..., [N/M_t]$) of the $p^{th}$ receiver branch is estimated with very small error, $\boldsymbol{\rho} - \boldsymbol{\rho}^* = \frac{1}{2}(\boldsymbol{\delta} - \boldsymbol{\delta}^*)$, that does not depend on the IQ-imbalance parameters and is caused solely due to CPE difference between two consecutive long training symbols. It is shown later in section IV that this error can be compensated for by applying a tracking algorithm. The channel vectors for the rest of the subcarriers can be estimated using interpolation or transform domain iterative techniques [19], [20]. To estimate the IQ-imbalance parameters, for $k = 0, ..., N - 2$, we define

$$\boldsymbol{\alpha}(k) = \mathbf{e}(k) - \mathbf{e}(k+1)$$
$$= \hat{\mathbf{h}}_p(k) - \hat{\mathbf{h}}_{p+1}(k+1) \tag{19}$$

$$\boldsymbol{\beta}(k) = \boldsymbol{\chi}_a(k) - \boldsymbol{\chi}_a(k+1)$$
$$= \mathbf{K}_1 \hat{\mathbf{h}}_p(k) - \mathbf{K}_1 \hat{\mathbf{h}}_{p+1}(k+1) \tag{20}$$

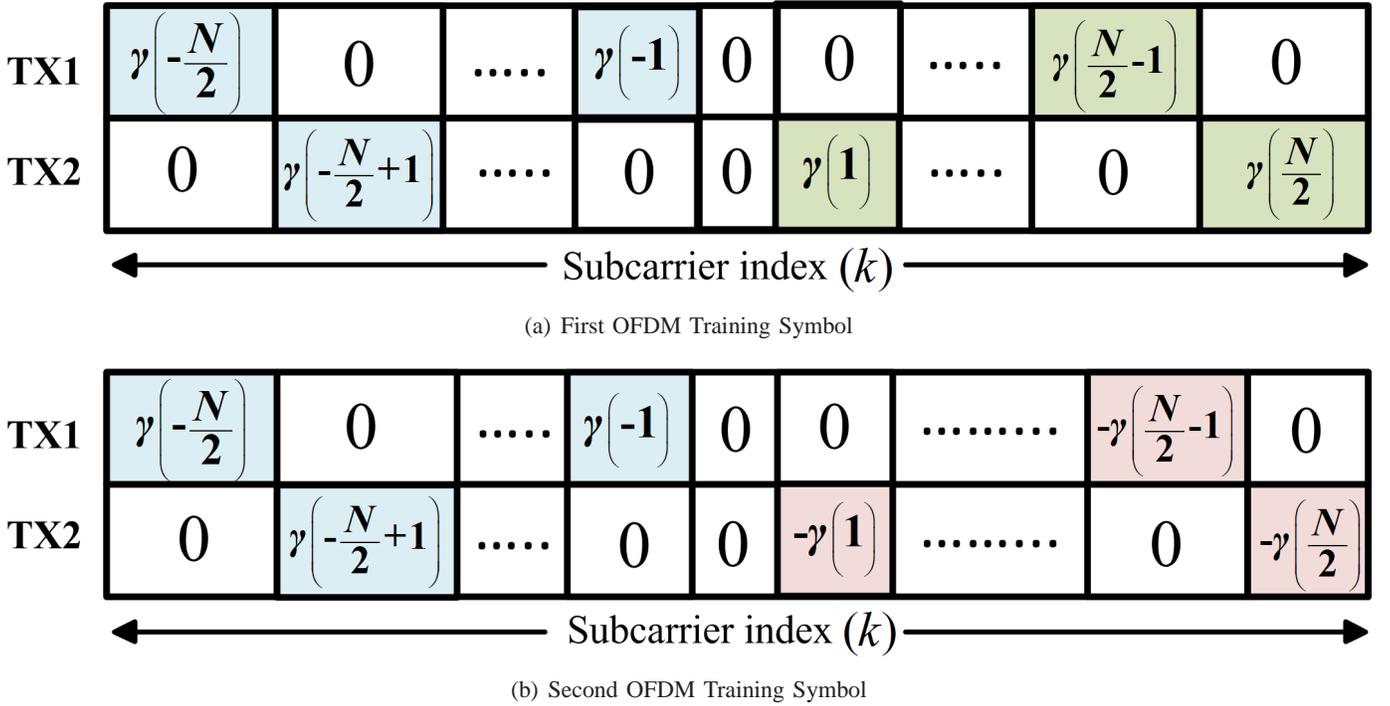

Fig. 3.  OFDM preamble structure for 2 transmit antennas

The phase and amplitude imbalance parameters are then calculated as

$$\tilde{\boldsymbol{\theta}}(k) = -\arg\{2(\boldsymbol{\beta}(k) \oslash \boldsymbol{\alpha}(k)) - \mathbf{1}_{M_r}\} \tag{21}$$
$$\tilde{\boldsymbol{\varepsilon}}(k) = \text{abs}\{2(\boldsymbol{\beta}(k) \oslash \boldsymbol{\alpha}(k)) - \mathbf{1}_{M_r}\} \tag{22}$$

where $\mathbf{1}_{M_r}$ is the $M_r \times 1$ all one vector and $\oslash$ represents element-wise division operation. The obtained imbalance parameters can be averaged over all subcarriers to get a better estimate. To further increase the accuracy of the estimates and because of the slow varying nature of IQ-imbalance parameters, these parameters can be averaged over several OFDM symbols (or frames).

## IV. Effective Channel Tracking and Data Detection

We assume quasi-static channel such that the channel response can be considered fixed within an OFDM frame. The CPE term, however, changes from one OFDM symbol to the other and needs to be estimated and compensated for each OFDM symbol. Assuming there are $r$ pilot subcarriers in $m^{th}$ OFDM symbol from $j^{th}$ transmit branch with pilot values $d_{m,j}(1), d_{m,j}(2), ..., d_{m,j}(r)$, it is possible to update the effective channel matrix estimates in (18) at data transmission stage as follows. The $M_r \times 1$ received pilot signals at $l^{th}$ and $-l^{th}$ subcarriers of the $m^{th}$ OFDM symbol where $l \in 1, 2, .., r$ can be written as (without considering ICI and noise)

$$\mathbf{x}_m(l) = \mathbf{K}_1 \boldsymbol{\Upsilon}_m \mathbf{y}_m(l) + \mathbf{K}_2 \boldsymbol{\Upsilon}_m^* \mathbf{y}_m^{\#}(l) \tag{23}$$
$$\mathbf{x}_m^{\#}(l) = \mathbf{K}_1^* \boldsymbol{\Upsilon}_m^* \mathbf{y}_m^{\#}(l) + \mathbf{K}_2^* \boldsymbol{\Upsilon}_m \mathbf{y}_m(l) \tag{24}$$

where, $\boldsymbol{\Upsilon}_m = diag\{\Upsilon_{m,1}, .., \Upsilon_{m,M_r}\}$ represents the CPE updating matrix such that $\Upsilon_{m,q} = \Theta_{m,q}(0)/\Theta_{pre,q}(0)$ and $\mathbf{K}_1 = diag(k_1^{(1)}, .., k_{M_r}^{(1)})$ and $\mathbf{K}_2 = diag(k_1^{(2)}, .., k_{M_r}^{(2)})$ are the IQ-imbalance matrices. We also define vector $\mathbf{y}_m(l)$ as $\mathbf{y}_m(l) = \{y_{m,1}(l), .., y_{m,M_r}(l)\}^T$ such that $\mathbf{y}_m(l) = \hat{\mathbf{H}}_{pre}(l)\mathbf{d}_m(l)$ where $\hat{\mathbf{H}}_{pre}(l)$ is the estimated channel matrix using training symbols as given in section III and $\mathbf{d}_m(l) = \{d_{m,1}(l), .., d_{m,M_t}(l)\}^T$ is the vector of transmitted pilots from different transmitter branch at $l^{th}$ subcarrier.

Algorithm 1 summarizes the procedure to compute the updating parameter matrix to compensate for varying CPE for each OFDM symbol. As the CPE is the same for all the subcarriers, it is possible to get a better estimate of

**Algorithm 1** Calculation of updating parameter matrix $\boldsymbol{\Upsilon}_m$

---

Input $\rightarrow \hat{\mathbf{K}}_1, \hat{\mathbf{K}}_2, \hat{\mathbf{H}}_{pre}(k), \boldsymbol{\Psi}, r, \mathbf{x}_m(l), \mathbf{x}_m^{\#}(l)$
Output $\rightarrow \boldsymbol{\Upsilon}_m$

- Initialize $\rightarrow q = 1$
  **while** $q \leq M_r$ **do**
  - Set $\mathbf{X}_{1,q} = \begin{bmatrix} k_q^{(1)} & k_q^{(1)} \\ k_q^{*(2)} & k_q^{*(2)} \end{bmatrix}$, $\mathbf{X}_{2,q} = \begin{bmatrix} k_q^{(1)} & -k_q^{(1)} \\ k_q^{*(2)} & -k_q^{*(2)} \end{bmatrix}$
    * Initialize $\rightarrow l = 1$
      **while** $l \leq r$ **do**
      1. Set $\boldsymbol{z}_q(l) = [x_{m,q}(l), x_{m,q}^{\#}(l)]^T$
      2. Compute $y_{m,q}(l)$ and $y_{m,q}^{\#}(l)$
      3. Compute $\hat{\mathbf{C}}_q(l) = y_{m,q}(l)\mathbf{X}_{1,q} + y_{m,q}^{\#}(l)\mathbf{X}_{2,q}$
      4. Compute $\mathbf{Q}_q(l) = (\hat{\mathbf{C}}_q^H(l)\hat{\mathbf{C}}_q(l) + \Psi_{q,q}\mathbf{I}_2)^{-1}\hat{\mathbf{C}}_q^H(l)$
      5. Compute $\hat{\boldsymbol{\varphi}}_q(l) = \mathbf{Q}_q(l)\mathbf{z}_q(l)$
      6. $l = l + 1$
  - Compute $\hat{\boldsymbol{\varphi}}_q = \frac{1}{r}\sum_{l=1}^{r}\hat{\boldsymbol{\varphi}}_q(l)$
  - Compute $\Upsilon_{m,q} = \hat{\varphi}_q(1) + j\hat{\varphi}_q(2)$
  - $q = q + 1$
- Set $\boldsymbol{\Upsilon}_m = diag\{\Upsilon_{m,q}\}_{q=1}^{M_r}$

---

the updating matrix by averaging it out over all pilot subcarriers. After finding the updating matrix, the transmitted data vector at the $k^{th}$ subcarrier, $\tilde{\mathbf{s}}_m(k)$, can be estimated using $\hat{\mathbf{H}}_m(k) = \boldsymbol{\Upsilon}_m \hat{\mathbf{H}}_{pre}(k)$, $\hat{\mathbf{K}}_1$, $\hat{\mathbf{K}}_2$ and solving

$$\min\|\hat{\mathbf{s}}_m(k) - \mathbf{A}(k)\hat{\mathbf{x}}_m(k)\|^2 \tag{25}$$

where $\hat{\mathbf{s}}_m(k) = \{\tilde{\mathbf{s}}_m(k), \tilde{\mathbf{s}}_m^{\#}(k)\}^T$ and $\hat{\mathbf{x}}_m(k) = \{\mathbf{x}_m(k), \mathbf{x}_m^{\#}(k)\}^T$ are the estimated and received signal vectors, respectively. For zero forcing (ZF) and minimum mean squared error (MMSE) based receivers

$$\mathbf{A}_{ZF}(k) = (\mathbf{W}^H(k)\mathbf{W}(k))^{-1}\mathbf{W}^H(k) \tag{26}$$
$$\mathbf{A}_{MMSE}(k) = (\mathbf{W}^H(k)\mathbf{W}(k) + \mathbf{R})^{-1}\mathbf{W}(k)^H \tag{27}$$

where

$$\mathbf{W}(k) = \begin{bmatrix} \hat{\mathbf{K}}_1\hat{\mathbf{H}}_m(k) & \hat{\mathbf{K}}_2\hat{\mathbf{H}}_m^{\#}(k) \\ \hat{\mathbf{K}}_2^*\hat{\mathbf{H}}_m(k) & \hat{\mathbf{K}}_1^*\hat{\mathbf{H}}_m^{\#}(k) \end{bmatrix} \tag{28}$$

and $\mathbf{R} = \boldsymbol{\Psi} \otimes \mathbf{I}_{M_r}$.

The complexity of our proposed algorithm can be analyzed in terms of the required complex additions and multiplications. The channel estimation in (16)-(18) only requires $3N$ complex additions and $2N$ complex multiplications. Also, the IQ-imbalance parameters estimation in (19)-(22) requires $2N$ complex additions and $2N$ complex multiplications. The estimation process does not involve any matrix inversion. The tracking algorithm involves inversion of a $2 \times 2$ matrix for $rM_r$ times. It also involves extra $13rM_r$ complex multiplications and $4rM_r$ complex additions. The scheme proposed in [12] is more complex than our proposed scheme since it involves higher number of complex multiplications. Note that in [12], the analysis is done only for SISO systems while the computational complexity increases exponentially for by the number of transmit/receive antennas.

## V. SIMULATION RESULTS

To validate the proposed scheme, a MIMO-OFDM system with spatial multiplexing transmission scheme [13] using 16-QAM modulation format is considered. The channel is assumed to be 7-tap multipath Rayleigh fading with exponential power delay profile for each transmit-receive link. We consider the 802.11a/n standard frame format in which each OFDM symbol is composed of 48 data, 4 pilot and 12 null subcarriers [16]. Cyclic prefix longer than the channel length is added at the beginning of each OFDM frame. The OFDM sampling time is considered to be $T_s = 0.05\mu s$. One short training symbol and two long training symbols are transmitted according to (10) and (11) in the beginning of each frame. One OFDM frame consists of 50 OFDM symbols (including training and data symbols) during which the channel is considered to be fixed. The IQ-imbalance parameters are chosen to be $5^o, 10\%$ for each receiver branch.

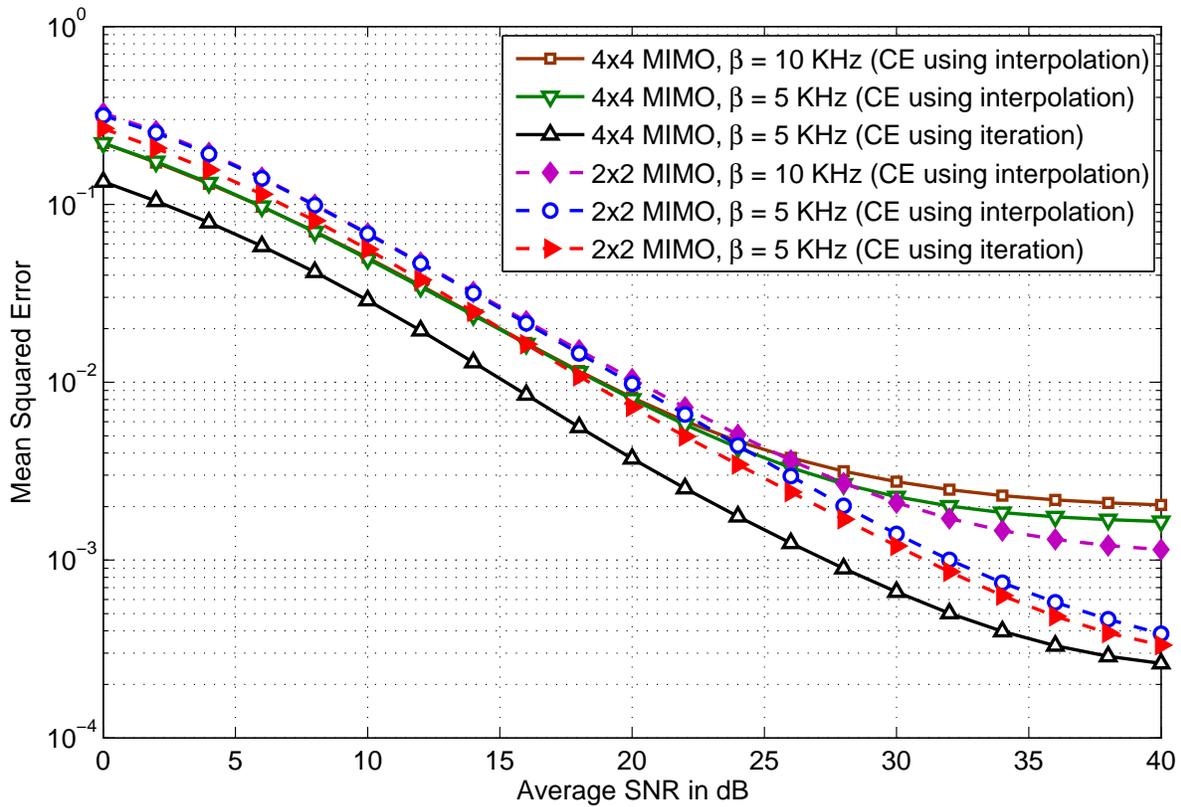

Fig. 4. MSE of channel estimation for $2 \times 2$ and $4 \times 4$ MIMO channels with IQ-imbalance of ($5^o$,10%) and different phase noise linewidths at each receiver branch

Fig. 4 shows the mean-squared error (MSE) of effective channel estimation (CE) for $2 \times 2$ and $4 \times 4$ MIMO channels with different phase noise linewidths, $\beta$. For channel estimation utilizing interpolation, spline interpolation technique is used and for iteration technique, the algorithm presented in [19] is used with 50 iterations. As shown, an increase in the linewidth of the phase noise increases MSE for both techniques as it introduces higher ICI. Also, there is little gain in terms of MSE in using iteration technique for $2 \times 2$ MIMO system compared to interpolation technique; however, this gain is larger in the case of $4 \times 4$ MIMO. The flooring effect in MSE at high SNRs is because of the error propagation caused due to the initial channel estimation error.

In Fig. 5, the MSE performance of the estimation of IQ-imbalance matrix ($\mathbf{K}_1$) using the metric $\mathbb{E}\{\|\hat{\mathbf{K}}_1 - \mathbf{K}_1\|^2\}$ is shown for a $2 \times 2$ MIMO channel with different phase noise linewidths. The estimated IQ-imbalance matrix, $\hat{\mathbf{K}}_1$, is calculated by averaging the phase and amplitude imbalances over all the subcarriers at the preamble stage and also over two consecutive OFDM frames. As shown, the IQ-imbalance parameters estimation performs very well even at high phase noise linewidths.

Figs. 6 and 7 show the performance in terms of bit error rate (BER) with phase noise linewidth of $\beta = 5$ KHz and IQ-imbalance of $5^o, 10\%$ at each receiver branch. As shown, the BER curves of the uncompensated and partially compensated systems have error floors at mid-range SNRs due to the combined effect of phase noise and IQ-imbalance. The performance of the proposed scheme, however, shows significant improvement compared to the ones that only compensate for IQ-imbalance or phase noise.

## VI. CONCLUSION

In this paper, a joint estimation and compensation scheme to mitigate IQ-imbalance and phase noise in MIMO-OFDM systems was presented. Assuming a suitable preamble structure, the channel and IQ-imbalance are initially

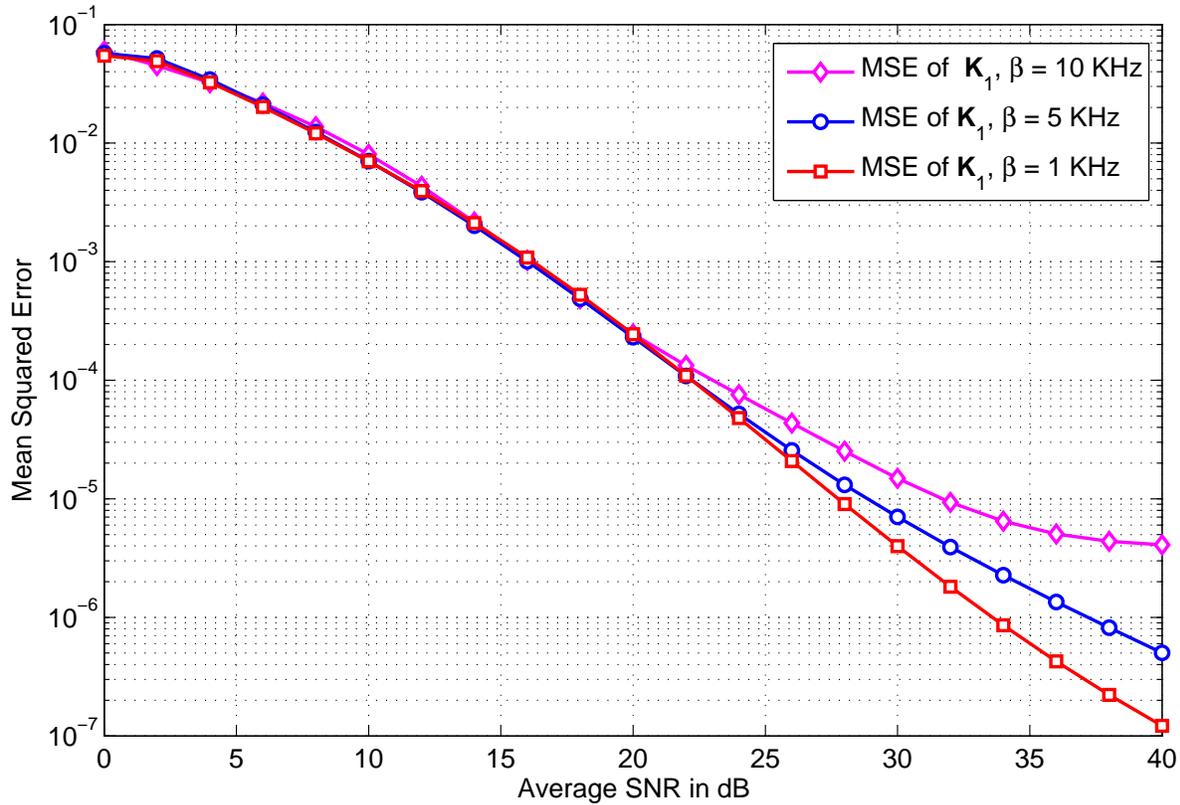

Fig. 5. MSE of IQ-imbalance matrix ($\mathbf{K}_1$) estimation for a $2 \times 2$ MIMO system with IQ-imbalance of ($5^{\circ}$,10%) and different phase noise linewidths at each receiver branch

estimated. Then, using pilot symbols in data stage of OFDM transmission, a method to update the effective channel estimates including CPE was discussed. The effective channel estimates can then be used to detect the transmitted data using detection schemes such as ZF and MMSE. The performance of the proposed scheme for spatial multiplexing in MIMO-OFDM systems was verified through simulation and it was shown that it provides significant performance improvement in terms of BER with reasonably low computational complexity.

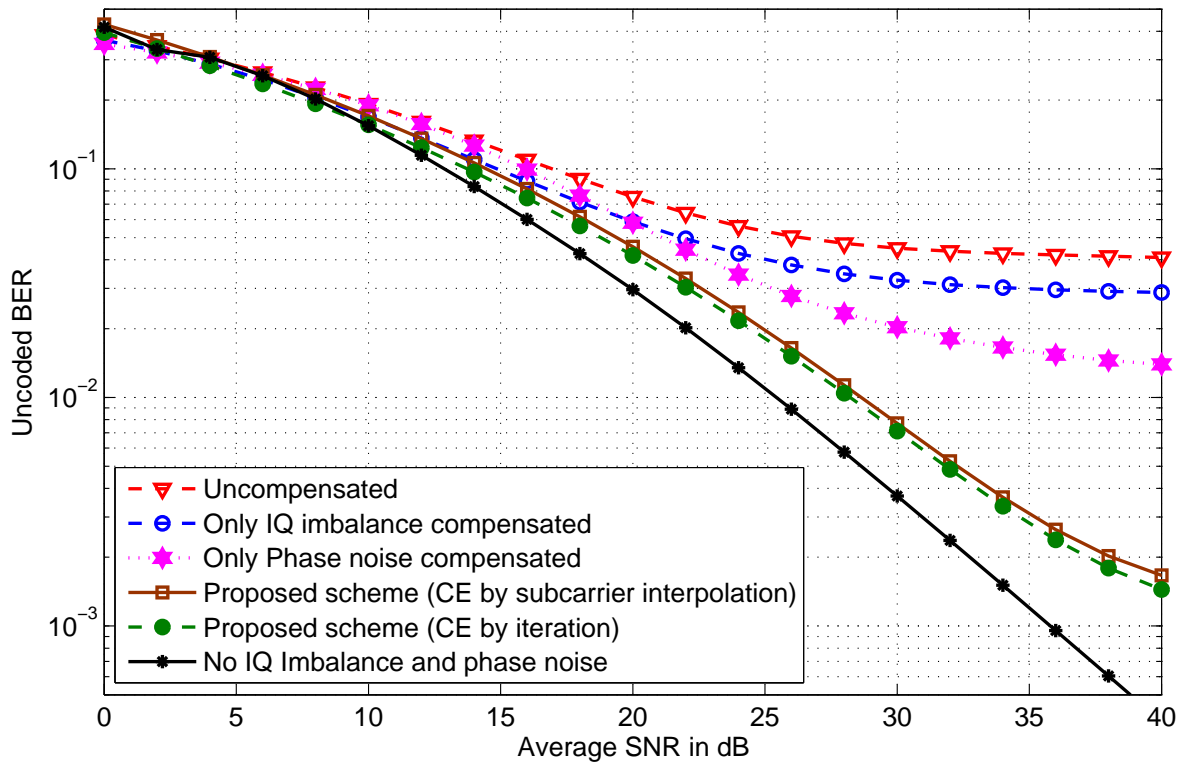

Fig. 6. BER performance of an uncoded 16-QAM, $2 \times 2$ MIMO-OFDM system with $\beta = 5$ KHz and IQ-imbalance of ($5^o$,10%) at each receiver branch

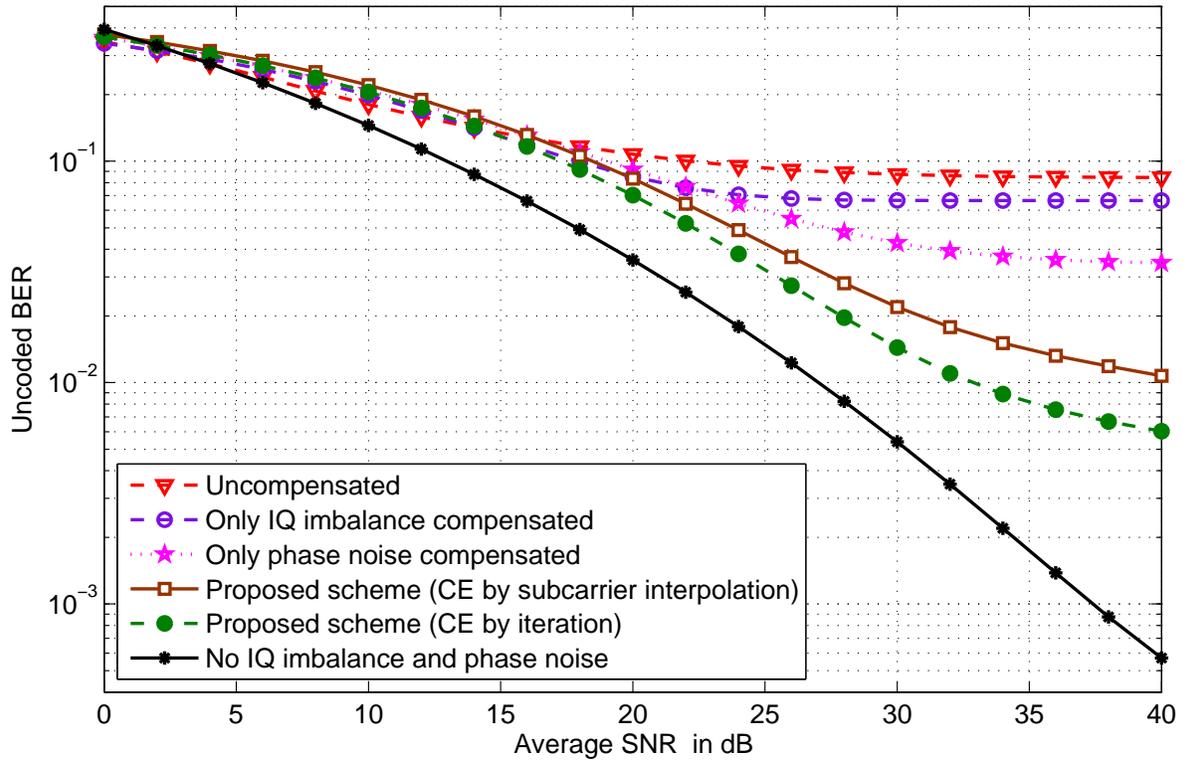

Fig. 7. BER performance of an uncoded 16-QAM, $4 \times 4$ MIMO-OFDM system with $\beta = 5$ KHz and IQ-imbalance of ($5^\circ$,10%) at each receiver branch

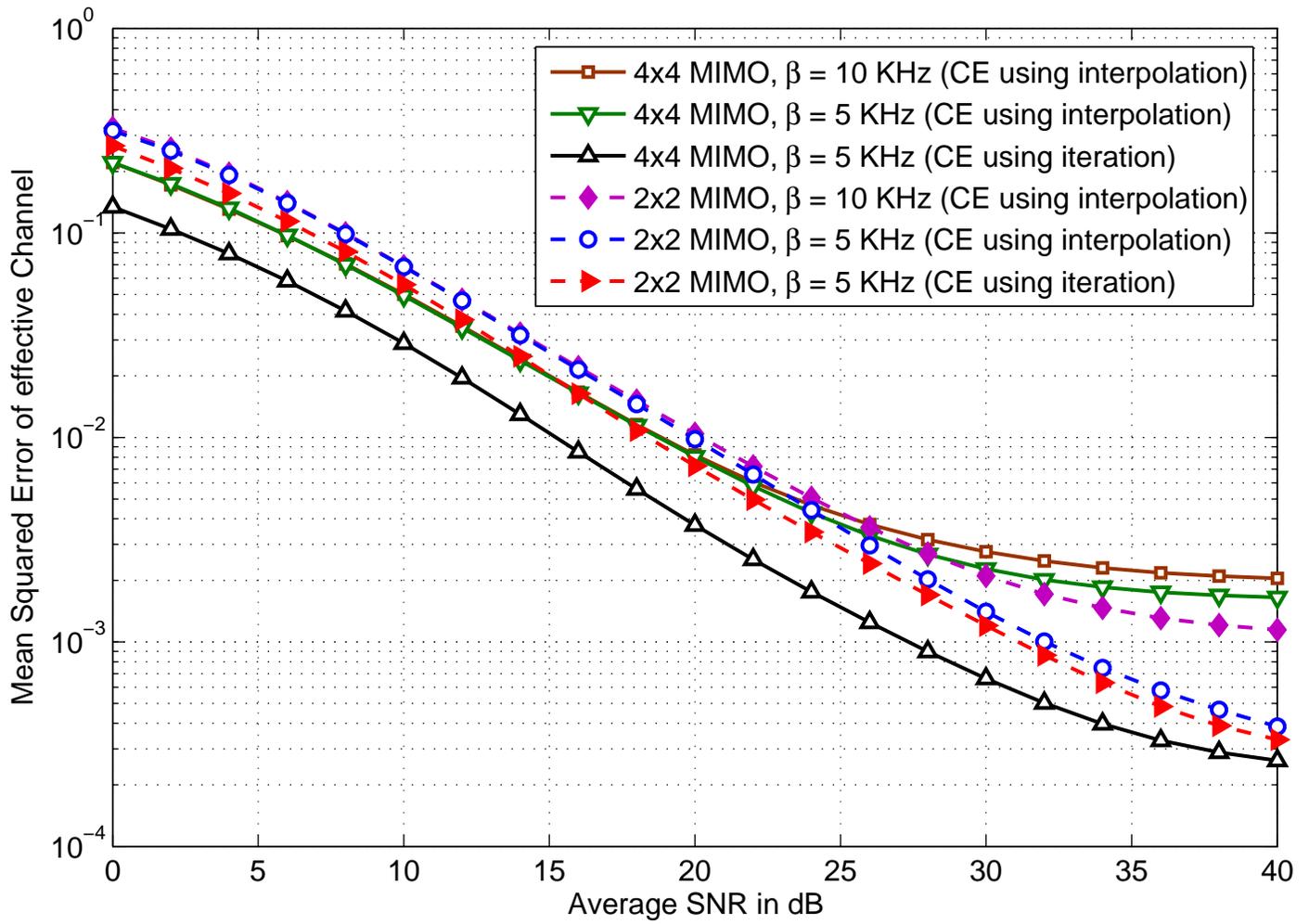